\documentstyle[aps,prb,psfig,preprint]{revtex}
\begin{document}
\draft
\title{
Collective Modes in High-Temperature Superconductors
}
\author{T. Dahm}
\address{Max-Planck-Institut f\"ur Physik komplexer Systeme,
N\"othnitzer Str. 38, D-01187 Dresden, Germany}
\author{D. Manske \cite{dmanske} and L. Tewordt}
\address{I. Institut f\"ur Theoretische Physik,
Universit\"at Hamburg, Jungiusstr. 9, D-20355 Hamburg,
Germany}
\date{\today}
\vspace{2ex}
\maketitle
\begin{abstract}
The role of collective modes in various experiments on the cuprates
is investigated. We calculate the neutron scattering, photoemission
(ARPES), and Raman scattering intensities below $T_c$
within the fluctuation-exchange (FLEX) approximation
for the two-dimensional Hubbard model.
It is shown that the large peak
in the dynamical spin susceptibility arises from a weakly damped
spin-density-wave collective mode. This gives rise to a dip between
the sharp low energy peak and the higher binding energy hump in
the ARPES spectrum. Furthermore, we show that the collective mode
of the amplitude fluctuation of the $d$-wave gap yields a broad peak
above the pair-breaking threshold in the $B_{1g}$ Raman spectrum.
\end{abstract}
\pacs{74.20.Mn, 74.25.Ha, 74.25.Jb, and 74.40.+k}
%
\narrowtext

A wide variety of collective modes has been observed in the
three phases of superfluid $^3$He \cite{lee}. These fluctuations
of the spin-triplet $p$-wave gap have been calculated from
coupled Bethe-Salpeter equations for the T-matrices in the
particle-particle and particle-hole channels \cite{einzel,schopohl}.
This method has also been used to investigate the collective
modes in hypothetical $p$-wave pairing superconductors like
Heavy Fermion superconductors \cite{monien}.
A detailed study of the collective modes for 3D $d$-wave
superconductors, including different order parameter symmetries,
has been made in Ref. \onlinecite{hirashima}.
Recently, pair fluctuations
and the associated Raman scattering intensity have been calculated
for a two-dimensional (2D) $d$-wave weak-coupling
superconductor \cite{wu}.

In this note we investigate the collective modes within the
fluctuation-exchange (FLEX) approximation for the
2D one-band Hubbard model and their relevance to neutron scattering,
photoemission, and Raman scattering experiments in high-$T_c$
cuprates. The FLEX-approximation for the
particle-hole channel \cite{bickers,dahm,tewordt} yields the dynamical
spin susceptibility, $\chi_s = \chi_{s0}(1-U\chi_{s0})^{-1}$, and
the charge susceptibility, $\chi_c = \chi_{c0}(1+U\chi_{c0})^{-1}$.
Here, $U$ is the on-site Coulomb repulsion, and
$\chi_{s0}({\bf q},\omega)$ and $\chi_{c0}({\bf q},\omega)$ are
the irreducible susceptibilities. The latter are calculated from
the {\it dressed} normal and anomalous Green's functions $G$ and $F$.
The corresponding normal self-energies, $\omega[1-Z({\bf k},\omega)]$
and $\xi({\bf k},\omega)$, and the $d$-wave gap function,
$\phi({\bf k},\omega)$, are determined self-consistently by the
Eliashberg equations with interactions given by
$(3/2)U^2\mbox{Im }\chi_s({\bf q},\omega)
\pm (1/2)U^2\mbox{Im }\chi_c({\bf q},\omega)$ (plus (minus) sign for
normal (anomalous) self-energy).

Below $T_c$ large peaks evolve in the spectral density
$\mbox{Im }\chi_s({\bf q},\omega)$, i.e., four distinct peaks at
wave vectors ${\bf q}$ near ${\bf Q}=(\pi,\pi)$ for next-nearest
neighbor hopping $t'=0$ \cite{dahm}, and a broad peak centered at
${\bf Q}$ for $t'=-0.45t$ (t is the near neighbor hopping energy)
\cite{tewordt}. These results are in qualitative agreement with neutron
scattering experiments on $\mbox{La}_{2-x}\mbox{Sr}_x\mbox{CuO}_4$
and $\mbox{YBa}_2\mbox{Cu}_3\mbox{O}_{7-\delta}$, respectively
\cite{rmignod}. Similar results have been obtained within the
$t$-$J$ model \cite{maki}. In Fig. \ref{fig1} we show
$\mbox{Im }\chi_s({\bf Q},\omega)$ for $U=3.6t$, $t'=-0.45t$, and
band filling $n=0.90$. One sees that a large peak evolves at
about $\omega_0=0.08t$ as $T$ decreases below $T_c=0.022t$. The
amplitude $\Delta_0$ of the $d_{x^2-y^2}$-wave gap rises much more
rapidly below $T_c$ than the BCS $d$-wave gap and reaches at our lowest
temperature $T=0.017t$ ($T/T_c = 0.77$) a value of about
$\Delta_0=0.1t$ (see Fig. \ref{fig3}). We find that the peak in
Fig. \ref{fig1} is due to a slightly damped collective mode
because the susceptibilty has a pole at $\omega_0$, more exactly,
$\mbox{Re }\chi_{s0}({\bf Q},\omega_0)-U^{-1}=0$, and the height
of the peak is large of the order of the quasiparticle lifetime
$1/\Gamma(\omega_0)$. Here, $\Gamma({\bf k},\omega)=\omega
\mbox{Im }Z({\bf k},\omega)/\mbox{Re }Z({\bf k},\omega)$ is the
quasiparticle scattering rate. Since this is decisive for the
observability of the collective modes in the cuprates we show in
Fig. \ref{fig2} the functions $\omega\mbox{Im }Z({\bf k},\omega)$
and $\mbox{Re }Z({\bf k},\omega)$ at the anti-node ${\bf k}_a$
and the node ${\bf k}_b$ of the gap on the Fermi line. One sees
from Fig. \ref{fig2} that for $T$ below $T_c$ the scattering rate
decreases dramatically for frequencies $\omega$ below the pair-breaking
threshold $2\Delta_0\simeq 0.2t$.

In order to understand somewhat better the origin of this
spin-density-wave collective mode we have calculated
$\chi_{s0}({\bf Q},\omega)$ in the weak-coupling limit. The
sums over Matsubara frequencies have been carried out with the
help of the methods developed for superfluid $^3$He \cite{schopohl}.
The effect of quasiparticle damping is taken into account by
carrying out the analytical continuation of this result from
$i\omega_m$ to $\omega+i\Gamma$. For a gap $\Delta({\bf k})=
(\Delta_0/2)(\cos k_x - \cos k_y)$ and a band
$\epsilon({\bf k})$ with $t'=0$
and chemical potential $\mu$ the summation over ${\bf k}$ in the
square Brillouin zone has been carried out numerically
in the following expression for $T=0$:
\begin{eqnarray}
\lefteqn{
\chi_{s0}({\bf Q},\omega) =
}\nonumber\\
& &
\sum_{k}\frac{E_kE_{k+Q} - \epsilon_k\epsilon_{k+Q}
- \Delta_k\Delta_{k+Q}}
{\left( E_k + E_{k+Q}\right)^2 - \left( \omega + i\Gamma\right)^2}
\,\frac{E_k + E_{k+Q}}{2E_kE_{k+Q}}
\label{eq:chi}
\quad .
\end{eqnarray}
Here, $E_k^2=\epsilon^2({\bf k}) + \Delta^2({\bf k})$.
Then we obtain a peak in the $\omega$-function $\mbox{Re }\chi_{s0}
({\bf Q},\omega)$ at the kinematical gap $\omega = 2 |\mu|$
\cite{bulut} whose height decreases with increasing $\Gamma$. The
approximate analytic result for $T=0$
is given by
\begin{displaymath}
\chi_{s0}({\bf Q},\omega) =
V_0^{-1} - N_F(z/1+z)^{1/2}\log\left[4\left(1+z\right)^{1/2}\right]
\quad ;
\end{displaymath}
\begin{eqnarray}
z & = & \left[ 4\mu^2 - \left(\omega + i\Gamma\right)^2\right]
/\left(2\Delta_0\right)^2
\quad ;
\nonumber\\
V_0^{-1} & = & N_F\log\left(2W/\Delta_0\right)
\label{z}
\quad .
\label{eq:chis0}
\end{eqnarray}
Here, $W=4t$ is the half bandwidth. The function
$\mbox{Re }\chi_{s0}({\bf Q},\omega)$ in Eq. (\ref{z}) rises
first with $\omega^2$ and then exhibits a peak at the kinematical
gap $2|\mu|$ whose height is about
$V_0^{-1} - (\pi/2)N_F(\Gamma/2\Delta_0)$. A low-frequency mode,
i.e., a zero of equation $\mbox{Re }\chi_{s0}({\bf Q},\omega)=1/U$,
is obtained only for a finite range of $U$ values which decreases with
increasing $\Gamma$.
For $t'=-0.45t$ a kinematical gap no longer exists and the effective
$|\mu|$ is nearly zero. Then the approximate analytical result for
the expression in Eq. (\ref{eq:chis0}) becomes equal to:
\begin{displaymath}
\chi_{s0}({\bf Q},\omega) = V_0^{-1} + \frac{1}{2}N_Fi
\frac{\bar{\omega}^2}{\left(\bar{\omega} + i\gamma\right)}
K(\bar{\omega} + i\gamma)
\qquad ;
\end{displaymath}
\begin{equation}
\bar{\omega} = \omega / 2\Delta_0
\qquad ; \qquad
\gamma = \Gamma / 2\Delta_0
\quad .
\label{eq:chisnull}
\end{equation}
Here, $K$ is the first elliptic integral. Now the peak of
$\mbox{Re }\chi_{s0}$ as a function of $\omega$ occurs approximately
at $\omega\simeq 2\Delta_0$. This result has been checked by carrying
out numerically the sum over ${\bf k}$ in Eq. (\ref{eq:chi}) for
$t'=-0.45t$ and for different amplitudes $\Delta_0$ and chemical
potentials $\mu$. In fact, we find that the function
$\mbox{Re }\chi_{s0}({\bf Q},\omega)$ exhibits a peak at about
$\omega\simeq 2\Delta_0$ whose height decreases as $\mu$ increases,
for example, from $-1.3$ (unrenomalized band filling $n_0=0.84$) to
$-0.8$ ($n_0=1.03$). A solution $\omega_0 < 2 \Delta_0$ of the
equation $\mbox{Re }\chi_{s0}({\bf Q},\omega_0)=1/U$ exists only for
a small range of $U$ values near $U\simeq 3t$. The strong-coupling
calculation yields a smaller resonance energy.
Another difference
in comparison to the weak-coupling result is the fact that the
self-consistent strong-coupling calculations yield a collective
mode for much higher values of $U$, for example, $U=6.8t$ in
Ref. \onlinecite{tewordt}. This shows how important it is to take
into account the feed-back effect of the self-energy on the
dynamical spin susceptibility $\chi_s$.
These strong renormalization effects might also be responsible
for the observed broadening and decrease of the resonance energy
of the neutron scattering peak in underdoped
$\mbox{YBa}_2\mbox{Cu}_3\mbox{O}_{6+x}$ which goes in proportion
to the decrease of $T_c$ or doping level \cite{fong}. In fact,
for decreasing doping $x=n-1$, or increassing chemical potential
$\mu$, the position and height of the function
$\mbox{Re }\chi_{s0}({\bf Q},\omega)$ decrease which means that the
position of the peak of $\mbox{Im }\chi_s$ is decreased and its
width is increased.

We show now that the spin-density-wave collective mode has a
large effect on the angle-resolved photoemission intensity (ARPES)
below $T_c$. In Fig. \ref{fig3}(a) we have plotted our results for
$N({\bf k},\omega)f(\omega)$ (where $N({\bf k},\omega)$ is the
quasiparticle spectral function and $f(\omega)$ is the Fermi function)
for several ${\bf k}$-vectors ranging from ${\bf k}=(\pi,0)$,
$(7\pi/8,0)$, $(13\pi/16,0)$, $\ldots$, down to $(0,0)$. The
parameter values are the same as in Figs. \ref{fig1} and \ref{fig2}.
For ${\bf k}$-vectors near $(\pi,0)$ we have a sharp low energy peak
followed by a dip and then a hump at higher energy. For ${\bf k}$ moving
from $(\pi,0)$ to $(0,0)$, the sharp peak remains first at about the
same position while the broad hump moves to higher binding energy.
In Fig. \ref{fig3}(b) we show the corresponding normal state
spectra at $T=0.023t$. One notices that the broad hump at higher
binding energy remains at the same position upon entering the
normal state, while the sharp peak and the dip feature disappear.
In the superconducting state along the nodal direction of the
$d$-wave order parameter we do not observe the dip feature as
we have shown previously \cite{tewordt}.
These results are
in qualitative agreement with the photoemission spectra of
$\mbox{Bi}_2\mbox{Sr}_2\mbox{CaCu}_3\mbox{O}_{8+\delta}$ (Bi 2212)
\cite{ding}. In this paper it was argued that the dip in the
spectrum stems from a step-like edge in the quasiparticle scattering
rate which arises from the interaction with a collective mode. This
scenario is confirmed by our results for the collective mode shown in
Fig. \ref{fig1} and by the scattering rate shown in Fig. \ref{fig2}.
We estimate from the edge of the peak in Fig. \ref{fig3} a gap amplitude
$\Delta_0\simeq 0.1t$ at $T/T_c=0.77$ and a spectral dip at binding
energy of about $2.3\Delta_0$ corresponding to a mode frequency
$\omega_0\simeq 1.3\Delta_0$ according to the estimates of Ref.
\onlinecite{ding}. However, here we have a discrepancy with regard to
the latter estimates because our mode frequency shown in Fig. \ref{fig1}
is much lower, i.e., $\omega_0\simeq 0.8\Delta_0$. We note that
we obtain also a dip in the density of states $N(\omega)$ below
the gap peak at negative $\omega$ values which agrees qualitatively
with STM measurements on Bi 2212 \cite{renner}. 

We want to mention
that higher order peaks in the photoemission spectra due to the
collective mode as have been observed for example in solid
hydrogen \cite{sawatzky} are not visible here. This is due to
the fact that in our case  the spin-density-wave collective mode 
is a damped mode and the strong quasi-particle damping rate
washes out higher order peaks. In addition, the self-energy
contains an average over momentum, further reducing this effect.
This is documented by the fact that the normal state spectrum
in Fig. \ref{fig3}(b), where there is no
collective mode present, is not much different from the
spectrum in the superconducting state in Fig. \ref{fig3}(a)
at higher binding energy.

We come now to the discussion of order parameter collective
modes in $d$-wave superconductors which can be calculated in
analogy to those in $p$-wave pairing superconductors \cite{monien}.
In general it can be said that the $d_{x^2-y^2}$-wave pairing component
in weak-coupling theory gives rise to the phase
fluctuation mode which is renormalized into a 2D plasmon \cite{wu},
and to the amplitude fluctuation mode of the $d$-wave gap.
For each additional
(weaker) pairing component, like an extended $s$-wave component,
one obtains an amplitude
(real) and a phase (imaginary) fluctuation mode.
Let us first consider the amplitude fluctuation
mode of the $d_{x^2-y^2}$-wave gap. We have calculated the
mode frequency $\omega_0$
from the weak-coupling expression
in Ref. \onlinecite{schopohl} for ${\bf q}=0$:
%
%
%
%
\begin{eqnarray}
\lefteqn{
\mbox{Re }\left[
\sum_k(\omega^2 - 4\Delta_k^2)\left[\cos(k_x)-\cos(k_y)\right]^2
\right.
}\nonumber\\
& \times &
\left.
\left[4E_k^2 - (\omega + i\Gamma)^2\right]^{-1}
\frac{\mbox{tanh}(E_k/2T)}{E_k}\right] = 0
\quad .
\label{nils}
\end{eqnarray}
By summing numerically over ${\bf k}$ in the square Brillouin zone
we obtain for $t'=0$ in $\epsilon({\bf k})$
and $T=0$ two solutions with frequencies
$\omega_0\simeq\sqrt{3}\Delta_0$ provided that the damping $\Gamma$ is
sufficiently large, namely, $\omega_0 < 3.5\Gamma$. For $t'=-0.45t$
and $T=0$ we obtain two solutions whose frequencies are somewhat
larger, $\omega_0\simeq 2\Delta_0$, where again the condition
$\omega_0 < 3.5\Gamma$ has to be satisfied. For a mode frequency
$\omega_0=2\Delta_0\simeq 0.2t$ at $T/T_c=0.77$ (see Fig. \ref{fig3})
one finds from Fig. \ref{fig2} a damping $\Gamma(\omega_0)\simeq 0.1t$
at the anti-node ${\bf k}_a$ which means that the condition
$\omega_0 < 3.5\Gamma$ is satisfied. In
Ref. \onlinecite{wu} a frequency $\omega_0=\sqrt{3}\Delta_0$
was obtained for the amplitude collective mode,
however, the coupling of this mode to the
charge fluctuations was neglected.
We find that the coupling
of this fluctuation in the particle-particle channel to the
charge fluctuation in the particle-hole channel yields
approximately the following
contribution $\chi_{fl}$ to the charge susceptibility $\chi_{c0}$
at $T=0$
(see Refs. \onlinecite{schopohl} and \onlinecite{monien}):
\begin{equation}
\chi_{fl}({\bf q}=0,\omega) =
2\left(\frac{N_F'}{N_F}\frac{1}{V_0}\right)^2\Delta_0^2
\,
\frac{1}{g(\omega)}
\quad ,
\label{chifl}
\end{equation}
\begin{eqnarray}
\lefteqn{
g(\omega) =
N_F\left[\frac{2}{3}\bar{\omega}^2 + \frac{4}{3}\gamma^2
- 1 -\frac{8}{3}i\bar{\omega}\gamma
\right.}
\nonumber\\
& + &
\left.
\gamma\left( 4\bar{\omega}^2 - 2\gamma^2 +
6i\bar{\omega}\gamma\right) \log\left( 4
\left[ 1 - \left( \bar{\omega} + i\gamma\right)^2\right]^{-1/2}
\right)\right]
;
\nonumber\\
&   &
\bar{\omega} = \omega / 2\Delta_0
\qquad ; \qquad
\gamma = \Gamma / 2\Delta_0
\quad.
\label{g}
\end{eqnarray}
Here, $N_F$ and $N_F'=dN_F/d\omega$ are the density of states and
its derivative at the Fermi energy $\omega=0$.
One notices from Eq. (\ref{g}) that in the limit $\gamma \to 0$
one obtains no valid solution of the equation
$\mbox{Re }g(\omega_0)=0$ because the solution $\bar{\omega_0}=
\sqrt{3/2}$ violates the condition that $\bar{\omega_0}\leq 1$.
However, for suffiently large values of $\Gamma$ ($\gamma \geq
\bar{\omega}/2$) one obtains a solution of the equation
$\mbox{Re }g(\omega_0)=0$ which satisfies the condition
$\bar{\omega_0}\leq 1$.
One sees from Fig. \ref{fig2} that
this condition is approximately satisfied for $\omega_0=2\Delta_0\simeq
0.2t$ because then one enters the pair-breaking continuum where
$\Gamma\sim\omega/2$ near the anti-node ${\bf k}_a$. Thus the
mode frequency is about $\omega_0=2\Delta_0$ for damping $\Gamma=\Delta_0$
in agreement with the numerical results.

We have calculated the resonance frequency of the exciton-like
$s$-wave mode of the order parameter which is caused by an
additional $s$-wave pairing component $|g_0|$ which is smaller than
the main $d$-wave pairing component $|\bar{g_2}|$ (see Ref.
\onlinecite{wu}).
The method of Refs. \onlinecite{schopohl} and \onlinecite{monien}
yields the following contribution $\chi_{exc}$ of this order
parameter fluctuation mode to the charge susceptibility $\chi_{c0}$
at $T=0$:
\begin{equation}
\chi_{exc}({\bf q}=0,\omega) =
-\left( N_F\omega\right)^2 \left[ g_{exc}(\omega)\right]^{-1}
\quad ,
\label{eq:chiexc}
\end{equation}
\begin{eqnarray}
g_{exc}(\omega)
& = &
\left( 1 - \frac{\bar{g_2}}{g_0}\right)
\sum_k \frac{1}{2E_k}
\nonumber\\
& + &
\frac{1}{2}\omega^2 \sum_k
\frac{\tanh(E_k/2T)}{E_k\left[ 4E_k^2 - \left(\omega
+ i\Gamma\right)^2\right]}
\quad .
\label{eq:gexc}
\end{eqnarray}
From Eq. (\ref{eq:gexc}) we obtain the following approximate
result:
\begin{equation}
g_{exc}(\omega) =
\left( 1 - \frac{\bar{g_2}}{g_0}\right)
\frac{1}{V_0} + N_F i \frac{\bar{\omega}^2}
{\left( \bar{\omega} + i\gamma\right)}
K(\bar{\omega} + i\gamma)
\label{eq:exc}
\end{equation}
where $V_0^{-1}$ is given by Eq. (\ref{eq:chis0}) and $\bar{\omega}$ and
$\gamma$ by Eq. (\ref{eq:chisnull}). We have carried out the summation
over ${\bf k}$ in Eq. (\ref{eq:gexc}) numerically and find in agreement
with Ref. \onlinecite{wu} that a solution of the equation in
$\mbox{Re }g_{exc}(\omega_0)=0$ at given $\Delta_0$ and $\Gamma$
exists only for very small values of the parameter $(\bar{g_2}/g_0)-1$
($\leq 0.1$). This means that the $s$-wave pairing coupling has to be
almost as strong as the $d$-wave pairing component which is quite
unrealistic. However, for increasing $\Gamma$ the resonance frequency
$\omega_0$ decreases and becomes much smaller than the pair-breaking
threshold $2\Delta_0$ for reasonably large scattering rates
($\Gamma/2\Delta_0 \sim 1/2$). This means that the contribution
$\mbox{Im }\chi_{exc}(\omega)$ of the exciton-like mode to Raman
scattering intensity with $B_{1g}$ polarization shows up as a small
peak below the pair-breaking threshold. Since the damping $\Gamma$
in the direction of the momentum of the antinode of the order
parameter rises rapidly with $\omega$ (see Fig. \ref{fig2}(a)) it
may be that this peak becomes observable for smaller values of the
ratio $g_0/\bar{g_2}$ of $s$-wave and $d$-wave pairing couplings than
those which have been obtained from weak-coupling
theory \cite{wu}.

In the weak-coupling limit it has been shown that vertex corrections due
to the $d$-wave pairing interaction together with
electron-electron scattering lead to good agreement with the $B_{1g}$
Raman data on YBCO \cite{manske}.
In Fig. \ref{fig4} we show our strong-coupling results for the Raman
response functions $\mbox{Im }\chi_{\gamma}({\bf q}=0,\omega)$
where $\gamma$ are the
vertices $\gamma=t[\cos(k_x)-\cos(k_y)]$ and $\gamma=-4t'\sin(k_x)
\sin(k_y)$ for $B_{1g}$ and $B_{2g}$
symmetry. One sees that for $B_{1g}$ symmetry a gap and a pair-breaking
threshold develop below $T_c$ with a threshold at about $0.15t\simeq (3/2)
\Delta_0$ at $T/T_c=0.77$ (see Fig. \ref{fig3}).
This means that the peak of
the order parameter collective mode at $\omega_0\simeq 2\Delta_0$ and
width $\Delta_0$ lies in the pair-breaking continuum.
The question arises whether or not the contribution of
$\mbox{Im }\chi_{fl}$
to the $B_{1g}$ Raman spectrum is sizeable because the coupling
strength proportional to $N_F'/N_F$ in Eq. (\ref{chifl}) arising from
particle-hole asymmetry is rather small.
However, in the strong-coupling calculation the coupling strength
of this mode to charge density given by
$T\sum_k\sum_nG({\bf k},i\omega_{n+m})F({\bf k},i\omega_n)$ is
much larger. The reason is that beside the term proportional to
$\epsilon({\bf k})$ yielding $N_F'/N_F$, one obtains additional terms
proportional to
the self-energy components $\mbox{Re }\xi({\bf k},\omega)$ and
$\mbox{Im }\xi({\bf k},\omega)$ which give relatively large
contributions. In addition, one obtains a contribution from the
imaginary part of the gap function, i.e. $\mbox{Im }\phi({\bf k},\omega)$.

In conclusion we can say that the spin-density-wave collective mode
below $T_c$ gives rise to large effects in the magnetic neutron
scattering and photoemission intensities and the tunneling density
of states. In order to explain the physical basis of our strong-coupling
results we have compared them with analytical expressions derived
from weak-coupling theory. This shows that the gap in the scattering
rate and the strong mass enhancement of the quasiparticles below
$T_c$ are decisive for the observability
of this mode.
On the other hand, the amplitude fluctuation
mode of the $d$-wave gap couples only weakly to the charge fluctuations
and yields a broad peak above the pair-breaking threshold in the
$B_{1g}$ Raman spectrum. This peak may be, at least partially,
responsible for the observed broadening above the pair-breaking
peak because
the coupling strength due to particle-hole asymmetry is enhanced
by strong-coupling self-energy effects.

Previous work on collective modes in high-$T_c$
superconductors \cite{wu,salkola,shen,dahm2,coffey}
has been restricted to weak-coupling and mean-field calculations.
The FLEX approach we use here, is a self-consistent and 
conserving approximation scheme, which goes well beyond 
mean-field. Especially the feed-back effect of the
one-particle properties on the collective modes in the
superconducting state is included self-consistently and
the importance of the quasiparticle damping becomes clear.
It is therefore a highly non-trivial and satisfactory
result, that the resonance in the spin-susceptibility, the 
step-like edge in the quasiparticle scattering rate, and
the dip-features in the ARPES and tunneling spectra can all be 
understood within one theory in a self-consistent fashion.
The self-consistent calculation also yields a larger coupling
strength of the $d$-wave amplitude mode to the charge
density and a lower resonance frequency of the $s$-wave 
exciton-like mode of the order parameter which makes it more
likely that these modes show up in the $B_{1g}$ Raman
scattering channel.

\acknowledgments
We acknowledge helpful discussions with D. Fay.
One of us (D.M.)
acknowledges financial support by the
Deutsche Forschungs\-gemeinschaft via the
Gra\-duier\-ten\-kolleg ''Physik nano\-strukturierter Festk\"orper''.
\begin{figure}
%
%
\caption{Spectral density of spin susceptibility at wave-vector
${\bf Q}=(\pi,\pi)$, $\mbox{Im }\chi_s({\bf Q},\omega)$, for
temperatures $T=0.023t$, $0.020t$, and $0.017t$ $(T_c=0.022t)$.
Here, $U=3.6t$ is the on-site Coulomb repulsion, $t$ the near
neighbor hopping energy, $t'=-0.45t$ the next-nearest neighbor hopping,
and $n=0.90$ the renormalized band filling.
}
\label{fig1}
\end{figure}
\begin{figure}
%
%
%
\caption{Quasiparticle scattering rate, $\Gamma({\bf k},\omega)=$
$\omega\mbox{Im }Z({\bf k},\omega)/\mbox{Re }Z({\bf k},\omega)$,
at anti-node ${\bf k}_a$ and node ${\bf k}_b$ of the
$d$-wave gap, in the normal state at $T=0.023t$ (solid lines),
and in the superconducting state at $T=0.017t$ $(T/T_c=0.77)$
(dashed lines). (a) $\omega\mbox{Im }Z({\bf k},\omega)$;
(b) mass enhancement $\mbox{Re }Z({\bf k},\omega)$.
}
\label{fig2}
\end{figure}
\begin{figure}
%
%
%
%
\caption{Photoemission intensity, $N({\bf k},\omega)f(\omega)$
(here $N$ is the quasiparticle spectral function and $f$ the
Fermi function) for ${\bf k}=(k\pi,0)$ where
$k=1$, $7/8$, $13/16$, $3/4$, $5/8$, $1/2$, $3/8$, $1/4$,
$1/8$, and $0$. (a) in the superconducting state at $T/T_c=0.77$.
The narrow peaks at low binding energy decrease
and vanish, and the binding energies of the broad humps increase
in the sequence of $k$ values. (b) in the normal state at $T=0.023t$.
Note, that the broad humps are at the same position as in the
superconducting state.}
\label{fig3}
\end{figure}
\begin{figure}
%
%
\caption{Raman spectra $\mbox{Im }\chi_{\gamma}({\bf q}=0,\omega)$
for $B_{1g}$ symmetry at $T=0.023t$ (solid line) and
$T/T_c=0.77$ (dashed line), and Raman spectrum for $B_{2g}$ symmetry
at $T/T_c=0.77$ (dotted line). $T_c=0.022t$.
}
\label{fig4}
\end{figure}
%
%

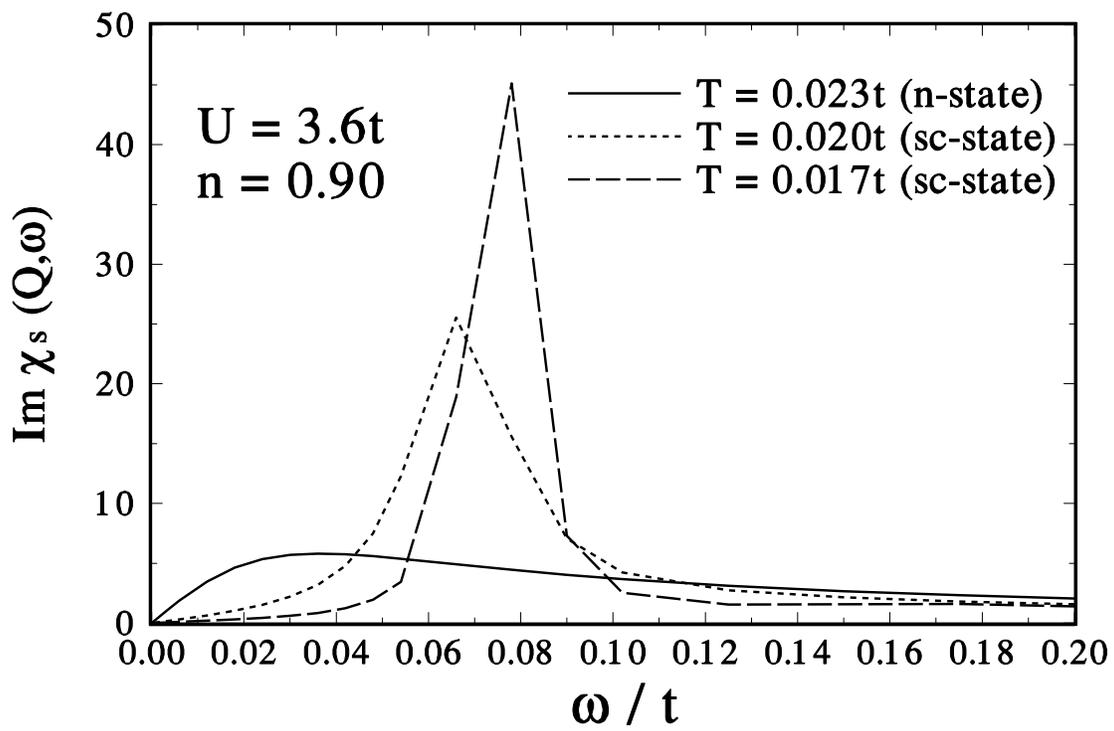
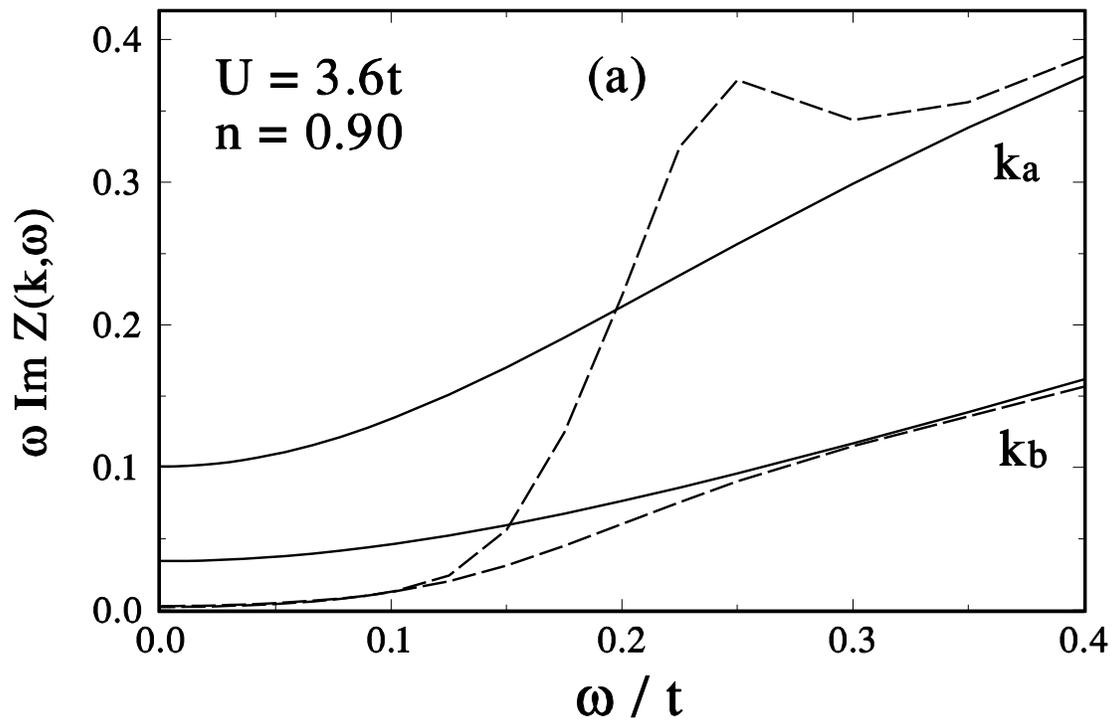
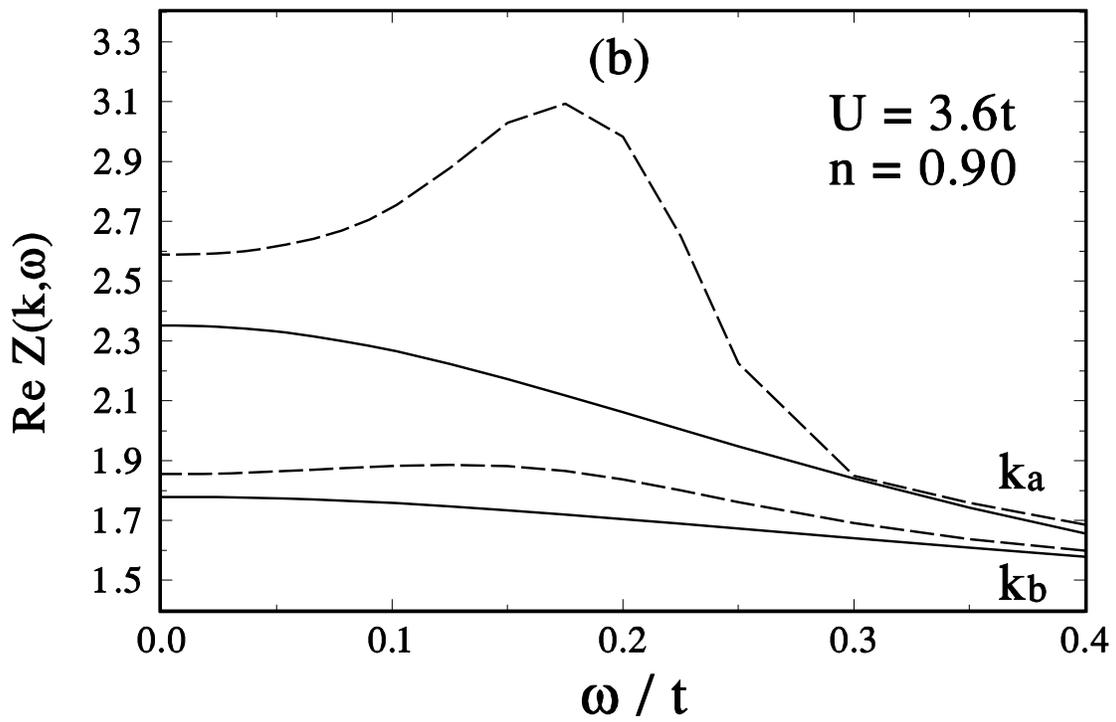
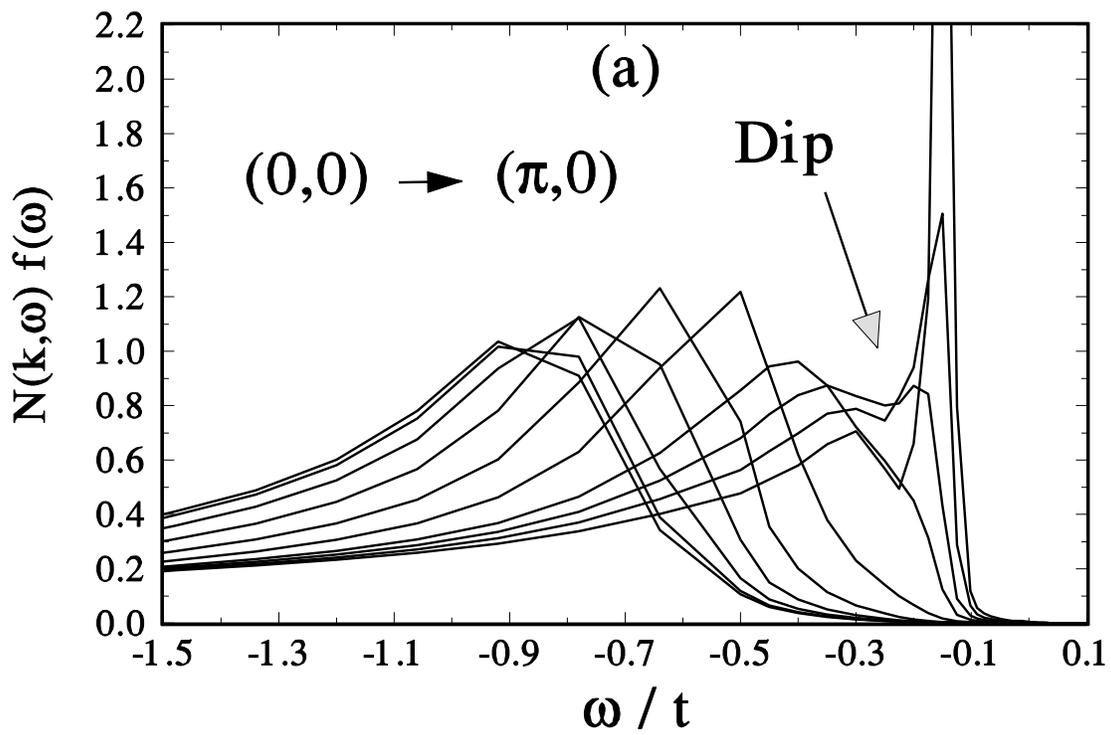
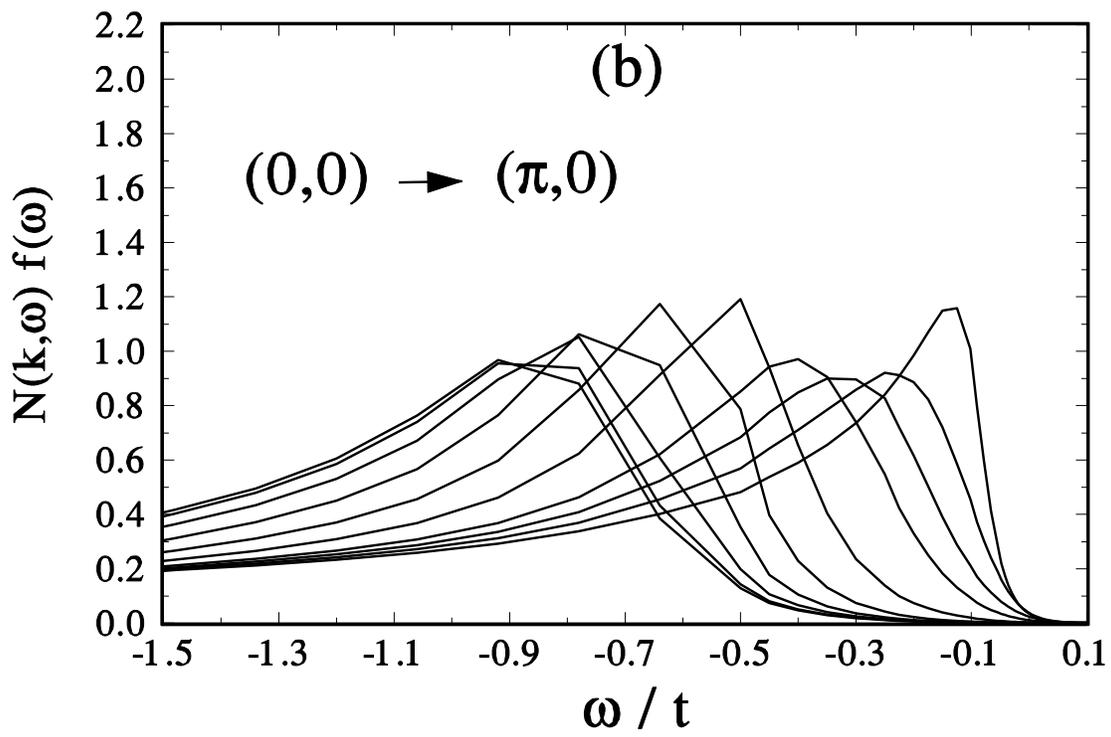
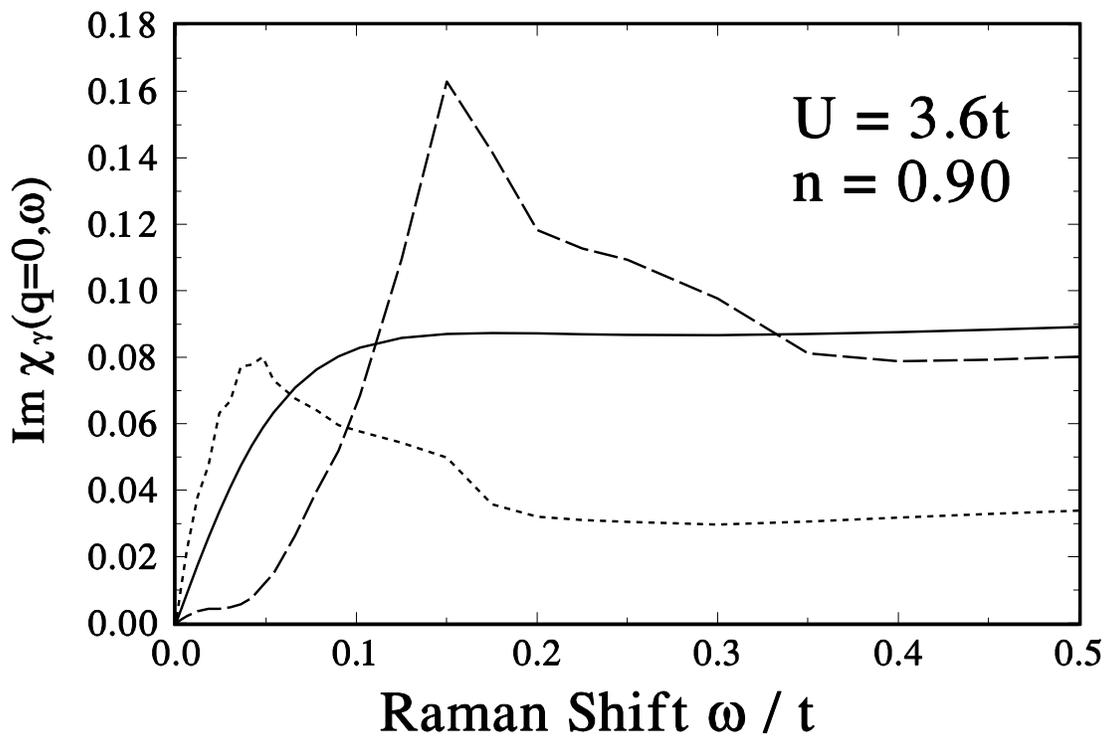
\end{document}